\documentclass{ws-procs9x6}

\begin{document}

\title{Messengers of the Extreme Universe}

\author{A. V. OLINTO}

\address{Department of Astronomy \& Astrophysics, Enrico Fermi
Institute, \\ Center for Cosmological Physics, University
of Chicago,\\ 5640 S.\ Ellis Avenue, Chicago, IL 60637, USA\\
E-mail: olinto@oddjob.uchicago.edu}

\maketitle

\abstracts{
We give a brief overview of the highest energy messengers: very high
energy photons,  neutrinos, and ultra-high energy cosmic rays. The
mysterious workings of the extreme universe should soon be unveil with
new observatories now under construction.Ê }

\section{Introduction}

The highest energy messengers arriving on Earth  bring information
from the most energetic events occuring in the universe since the Big
Bang. These extreme environments cannot be reproduced in terrestrial
laboratories. Instead, physics at these extremes is studied using
telescopes and detectors that span many orders of magnitude in energy
(from 10$^{9}$ eV photons to 10$^{20}$ eV cosmic rays).
Sources range from Supermassive black holes (masses $\sim 10^9 M_{\odot}$)
in the centers of galaxies (known as active galactic nuclei or AGN) to
yet to be discovered sources of the highest energy cosmic rays.  These
astrophysical environments not only call for applications of  physics to
extreme cosmic systems but they also challenge fundamental physics beyond
what is currently known.

\section{High-Energy Gamma-rays}

The most important messengers in the history of astronomy and
astrophysics are photons. Visible photons started astronomy and photons
are still the clearest messengers of most phenomena in the universe
since photons are neutral, massless, abundantly
produced, and easily detected. They move the fastest and in the absence of
strong gravitational fields, they point straight back to their sources. In
contrast, cosmic protons and electrons are charged and usually diffuse in
cosmic magnetic fields. Electrons and positrons lose energy fast
while unstable particles, such as neutrons, decay on their way to Earth.
Neutrinos have some of the advantages of photons, they are neutral and
have very small masses, but they only interact through the weak
interactions, making their detection especially challenging.  
Photons are easily detected and easily produced by cosmic events that
accelerate charged particles. We observe cosmic photons from radio
wavelengths ($\sim 10^{-7}$ eV) to the highest energy photons observed
thus far, gamma-rays of tens of TeVs (1 TeV = 10$^{12}$ eV). Here we
focus on gamma-rays, i.e, photons above 10$^5$ eV.

Gamma-ray astronomy was born once rockets and balloons reached above the
absorbing atmosphere of the Earth. By 1961, the Explorer-XI satellite
had detected a few gamma-rays consistent with a cosmic gamma-ray
background. By the 1970s, SAS-2 and COS-B satellites produced maps of
the sky in gamma-rays where the Galactic gamma-ray emission, the
gamma-ray background, and the presence of a few point sources were
identified.  The most surprising discovery in early gamma-ray astronomy
was made by the defense Vela satellite series. These were designed to
monitor compliance with the 1963 Limited Test Ban Treaty by detecting
flashes of gamma-rays from man-made nuclear explosions. Instead, Vela
4A and 4B in 1967, followed by Vela 5A and 5B in 1969, discovered many
flashes of gamma-rays coming not from the direction of the Earth but
from space. These gamma-ray bursts (GRBs) are bright flashes of hard
X-rays and gamma rays that last fractions of a second to minutes during
which they dominate the gamma-ray sky. Afterglows in lower energy
wavelengths have shown that these bursts originate in some highly
energetic explosions in galaxies throughtout the observable universe.

Great progress in observations of cosmic gamma-ray sources and gamma-ray
bursts was achieved with the launch of the Compton Gamma-Ray Observatory
(CGRO), on April 4th, 1991. The Burst And Transient Source Experiment
(BATSE) aboard CGRO  recorded over 2700 gamma-ray bursts that showed a
clearly isotropic distribution. The highest energy instrument on CGRO,
the Energetic Gamma Ray Experiment Telescope (EGRET), surveyed the
whole sky survey from 20 MeV to 30 GeV and detected 66 blazar AGN, 27
lower-confidence AGN, 7 pulsars, many gamma-ray bursts, and 170
unidentified sources  (see Grenier\cite{grenier} in these proceedings
for more on gamma-ray sources). In addition, EGRET set a limit on
the extragalactic diffuse gamma-ray background that has constrained
many models of ultra-high energy phenomena. Compton re-entered the
Earth's atmosphere on June 4, 2000.

The next generation of gamma-ray satellites will have
improved sensitivity at higher energies. The Gamma-Ray Large Area Space
Telescope (GLAST) will have a large area ($\sim$ 1 m$^2$) and will
observed gamma-rays from 10 MeV to above $\sim$ 100 GeV with excellent
energy and angular resolutions. GLAST is expected to be launched in
2006. Before GLAST, the Astro-rivelatore Gamma a Immagini Leggero
(AGILE) will cover the energy range from 30 MeV
to 50 GeV while Astro-E2 should be launched in 2005 with a range from
0.4 keV to 700 keV. AGILE will have similar sensitivity to EGRET with a
larger field of view and much better angular resolution.

Cosmic gamma-rays of energies above 100 GeV are hard to detect from
space but can be studied on the ground. High-energy gamma-rays
entering the atmosphere interact with the atmospheric particles and
develop a cascade of particles known as an extensive air-shower. The
electromagnetic cascade can be imaged by Atmospheric Cherenkov
Telescopes (ACTs) such as the Whipple 10 m telescope in Arizona, the
CAT telescope in the Pyrenees, the Cangaroo telescope in Australia, and
the HEGRA telescope in the Canary Islands. ACTs detect the Cherenkov
radiation generated by the electromagnetic particles in the shower from
gamma-ray primaries with energies $\sim$ 100 GeV or higher. For lower
thresholds, the STACEE and CELESTE experiments utilize large arrays of
heliostats at solar power plants to reach just
below 100 GeV where overlap with satellite data can occur. For very
high energies, the shower is detectable at the ground by Extensive Air
Shower (EAS) arrays like CASA and CYGNUS. These arrays sample the
showers sparsely and have $\sim$ 100 TeV thresholds. To lower the
threshold of EAS arrays to $\sim$ 1 TeV, the Milagro experiment has a
large active area employing water as the detecting medium and an array
of photomultiplier tubes. The Tibet array can also reach lower
thresholds by being located at  high altitude. To
observe a wide range of gamma-ray energies, a variety of ground
detectors need to complement the lower energy satellite observations.

Shower pattern imaging techniques have improved substantially in modern
ACTs increasing significantly the ability to reject cosmic ray
background showers. Imaging ACTs were the first telescopes to
observe TeV sources (see Aharonian\cite{aharonian} for more on TeV
sources). A new generation of ground-based instruments is under
development such as HESS, MAGIC, CANGAROO-III, and VERITAS. These ACTs
will reach the highest gamma-ray energies and will rival space-based
instruments in sensitivity. Satellites provide a platform for all sky
surveys and background studies while ground telescopes are best suited
to observe known point sources. These two approaches are highly
complementary in energy range and type of source monitoring and
together will help understand some of the highest energy environments
in the universe.

Gamma-ray bursts show that the universe is quite transparent to 100 MeV
gamma-rays, but the universe is liekly to be opaque to very high energy
gamma-rays. Pair production in extragalactic diffuse backgrounds are
likely to degrade gamma-rays above $\sim$ 10 TeV to $\sim$10$^{19}$ eV
producing a natural limit to the reach of very high energy gamma-ray
observatories. It is clear that the CMB stongly supresses high redshift
PeV ($10^{15}$ eV) gamma-rays, but the infrared and radio backgrounds,
that influence the TeV and EeV gamma-ray propagation respectively, are
less well determined. These less known backgrounds can be studied by
contrasting high redshift to low redshift gamma-ray emission.  For
instance, EGRET observed many blazars up to $\sim$ 30 GeV which have
not been detected at $\sim$ TeV energies. In principle, this could be
due to a cutoff in the blazar spectrum, but since the few TeV blazars
that have been detected are all at relatively low redshifts ($z
\sim $ 0.03 to 0.1), it is more likely that the lack of observations be
due to an energy dependent gamma-ray horizon. The lack of TeV emission
from powerful high redshift blazars gives an estimate of the infrared
background. Finally, if high redshift gamma-ray sources have spectra
that extend well beyond TeV energies, the effect of the diffuse
backgrounds is to shift the high energy radiation to the EGRET energy
range. Thus, EGRET limits on the extragalactic gamma-ray background are
strongly constraining to models where ultra high energy gamma-rays are
produce throughout the universe.

\section{High-Energy Neutrinos}

Neutrino astronomy was born in the mid 1960s with Davis' chlorine
experiment detecting neutrinos from the $^8$B beta decay in the
core of the Sun. The discrepancies between the observed flux and the
theoretical predictions led to the well known solar neutrino problem.
A number of experiments pointed to the solution of this problem, most
recently the Sudbury Neutrino observatory (SNO), which has the ability to
detect all three flavors of neutrinos, was able to show that the
total flux of the three neutrino flavors matches the theoretical
predictions and the deficit of electron neutrinos from the Sun is due to
their oscillations most likely into tau neutrinos. Another neutrino
flavor oscillation was shown by SuperKamiokande (SuperK) to explain the 
atmospheric neutrino problem: a deficit of muon neutrinos relative to
electron neutrinos produced by cosmic ray interactions in the Earth's
atmosphere.  The solution to these problems is now known to be a great
new discovery in fundamental physics: neutrinos have masses and
oscillate between different flavors. (For the detection of cosmic
neutrinos, Davis and Koshiba from SuperK were awarded the 2002 Nobel
prize in physics.)

Another great discovery involving neutrinos was the confirmation of the
standard scenario of core-collapse supernova by the detection of the
first extrasolar neutrinos in 1987. Kamiokande detected 12 neutrinos from
Supernova 1987a while IMB detected 9 events. Even though only a few
neutrinos were measured from SN1987a, this detection led to great
improvement in our understanding of core collapse supernovae. If a
future supernova is observed in our own Galaxy, the richness of
information carried by the neutrinos would be unparalleled.

Although the observed neutrinos from the SN87a are relatively low energy
neutrinos, higher energy neutrinos are also expected to be emitted by
supernovae explosions and by most high-energy astrophysical systems in
the universe. The advantage of searching for very high-energy neutrinos
from astrophysical sources is that, unlike gamma-rays, neutrinos have
no horizon due to interactions; they can traverse the observable
universe unimpeded. In particular, gamma-ray bursts and AGN should
produce observable fluxes of neutrinos from nearby to very high
redshifts. The estimated fluxes indicate a minimum scale of km$^3$ or 1
billion tons of detecting material. There are two approaches being
taken for neutrinos around 1 PeV (10$^{15}$ eV) in energy: km$^3$ water
detectors (ANTARES, NEMO, NESTOR) and km$^3$ ice detectors (IceCube).
These detectors use the Earth as a filter of background particles and
observe upward going muons generated by neutrinos as they traverse the
Earth. High-energy muons emit Cherenkov radiation in water or ice,
which is then observed by an array of photomultiplier tubes. Prototype
detectors in ice (AMANDA) and water (Lake Baikal) have already observed
many events consistent with atmospheric neutrinos. (See
Halzen\cite{halzen} in these proceedings for more on high energy
neutrinos.) 

Another source of even higher energy neutrinos are extragalactic
ultra-high energy cosmic rays. As ultra-high energy cosmic rays traverse
the universe, they interact with the cosmic microwave background
radiation and generate pions. Pions decay generating neutrinos with
energies starting $\sim 10^{17}$ eV and higher. The new generation of
ultra-high energy cosmic ray detectors, such as the Auger Project and
the EUSO satellite experiment, will be able to study ultra-high energy
neutrinos through horizontal and Earth skimming showers. In addition,
new techniques for ultra-high energy neutrino observations using radio
signals in ice (e.g., RICE and ANITA) or salt mines (e.g., SALSA) are now
under development. Over the next decade, new high-energy neutrino
detectors will open a new window of high-energy astrophysics. (See
Klages\cite{klages}, Blasi\cite{blasi}, and Berezinsky\cite{berezinsky}
in these proceedings for more on high energy neutrinos from
ultra-high energy cosmic ray sources.)

\section{Cosmic-Rays Ð the highest energy messengers}

Observations of extensive air showers have detected the highest energy
particles ever observed: cosmic rays with macroscopic energies around
tens of Joules or 10$^{20}$ eV. Cosmic rays with energies up to 10$^{14}$ eV
can be detected directly with balloon and space experiments while
above this energy the flux is too low for space-based detectors and
cosmic rays are studied through their air shower development.  The cosmic ray spectrum from
$\sim$ 10$^{9}$ eV to 10$^{14}$ eV  is
well described by a power law of spectral index $\sim$ 2.7.  At higher
energies the spectrum steepens into a spectral index $\sim$ 3 and the
transition region is called the knee. At the highest energies the
spectrum seems to change again but the exact features are still
unclear. 

Composition studies at low energies show a diffusive history of cosmic
ray nuclei through the Galaxy. At lower energies, cosmic rays are
dominated by light nuclei (protons and helium) while at the knee the
composition seems to become heavier (see Haungs\cite{haungs} in these
proceedings for recent results on composition from Kaskade). Galactic cosmic ray
propagation is diffusive in magnetic fields with a rigidity
dependent escape probability. The knee is thought to  represent the
transition from confined trajectories to trajectories that escape the
Galaxy and thus the change in spectral index. The origin and
propagation history of cosmic rays are still unclear with the leading
theory being shock acceleration in supernova remnant as proposed by
Enrico Fermi in 1949. A clear confirmation of this picture is still
lacking but can be achieved by gamma-ray detectors sensitive to
gamma-rays from pion decay around the acceleration sites. 

Fermi acceleration in supernova remnants may be responsible for
accelerating cosmic rays below the knee, but more powerful sources seem to
be required for the higher energy events. In addition, as the energy of
the primary cosmic ray increases, the effect of the Galactic magnetic
field in the particle trajectory decreases. As cosmic rays reach ~ 10$^{19}$
eV and above, trajectories should start to point back to cosmic ray
sources, i.e., cosmic ray astronomy should become possible. Instead,
observations show an isotropic distribution of arrival directions up to
the highest energies observed thus far. With no indication of the
Galactic plane or other nearby structures, the isotropy argues for an
extra-galactic origin for the highest energy sources of cosmic ray
protons or much stronger than expected magnetic fields in the halo or
our Galaxy.

Ultra-high energy cosmic rays are detected mainly via two techniques:
ground arrays of scintillators or water Cherenkov tanks and fluorescence
detectors. Ground arrays sample the extensive air shower as it reaches the
ground while fluorescence observatories detect the fluorescence of
nitrogen molecules in the atmosphere as the shower develops above the
ground. This second technique works best in clear moonless nights and in a stereo mode, when two detectors separated by many kilometers can observe the same shower
from two different angles determining the shower axis. The leading experiments using each of these techniques, the Akeno Giant Air Shower Array (AGASA) and the High Resolution Fly's Eye (HiRes), are now probing the  predicted Greisen-Zatsepin-Kuzmin (GZK) cutoff\cite{gzk}  (see Nagano\cite{nagano} in these proceedings for more on these experiments).
A cutoff should be present if the ultra-high energy particles are extragalactic protons,  because protons above $7 \times 10^{19}$ eV produce pions on their way to Earth through interactions with the cosmic microwave background. The energy lost due to the photo-pion production generates a feature in the spectrum, the GZK feature.
Thus, in addition to the extraordinary energy requirements for
astrophysical sources to accelerate protons to 10$^{20}$ eV, the photopion
threshold reaction was expected to cause a natural high-energy limit to the
cosmic ray spectrum known as the GZK cutoff.

The first definitive detection of an ultra high energy cosmic ray was recorded by
the Fly's Eye experiment in 1991: a 3.2 $\times 10^{20}$ eV event. This
event triggered considerable interest on the origin and nature of these
particles that indicate the  lack of the predicted GZK cutoff. 
Following this event, AGASA has reported 11 events above the GZK cutoff which argues for the absence of the cutoff. Together with the isotropy of arrival directions these events pose a challenge to models for the origin of UHECRs.
However, the HiRes collaboration reported a suppression of the UHECR flux
at high energies based on an analysis of their monocular data. With the
presently available low statistics of observed extremely high energy
events, the GZK feature associated with the photopion production is still
dominated by fluctuations\cite{demarco} an a new generation of
experiments can resolve this issue (see Blasi\cite{blasi} and
Berezinsky\cite{berezinsky} in these proceedings). 

The hybrid detector of the Auger Project (see Klages\cite{klages}) should settle the present disparity between AGASA and HiRes. The Pierre Auger Project is now under construction
in Argentina and a second site should be constructed in the northern
hemisphere. The Argentinean site will have a 3000 km$^2$ array of water
Cherenkov tanks together with four fluorescence detectors overlooking the
array. The use of both surface array and fluorescence technique will help
resolve the apparent inconsistency in the spectrum between AGASA and
HiRes by measuring many more events and cross calibrating the two
techniques. It can also determine the composition by measuring the depth
of shower maximum and the ground footprint of the same shower. 
The solution to the ultra-high energy
cosmic rays mystery will come with the next generation of experiments,
which are under construction such as Auger or in the planning stages such
as the space-based EUSO and OWL observatories. These space fluorescence
detectors can monitor extensive air showers from above the atmosphere on
very large volumes.

\section{The Future of High energy astrophysics} 

The future of high-energy astrophysics looks extremely promising. The
combination of space missions that span most wavelengths of detectable
photons and ground-based detectors that reach thousands of km$^2$ will
bring the many puzzles involved in the production and propagation of the
highest energy gamma-rays, cosmic-rays, and neutrinos into a new era. 

After the success of CGRO, the next large gamma-ray space-based
observatory now under construction is GLAST. GLAST's new technology
includes Si trackers, CsI calorimeters, and anti-coincidence veto and is
expected to be launch in 2006. At the higher energies, ground based
gamma-ray telescope arrays are now being built such as the Tibet Array,
HESS (started in 2002), VERITAS and Cangaroo III (now under
construction). At intermediate energies, MAGIC follows STACEE and CELESTE
in the overlap with the satellite energy range. 

High-energy neutrino observatories will be able to reach the necessary
km$^3$ scale in ice (IceCube) and in water following ANTARES, NESTOR and
NEMOS. These observatories will open a new window into the workings of
the high-energy systems in the universe. At the highest energies,
neutrinos should be detectable by the next generation ultra-high energy
cosmic ray detectors such as Auger, EUSO, and OWL and the next generation
radio detectors such as SuperRICE and ANITA.

In ultra-high energy cosmic rays, the future promises to unveil the
origin and nature of these mysterious particles. The present state of
observations is particularly puzzling and the necessary experiments to
resolve these puzzles will be operating in the very near future. The
Pierre Auger project should be fully operational in the Southern site in 2005 and
the Northern site shortly after. Once Auger reaches past 10$^{20}$ eV, further
exploration into higher energies can be achieved by space observatories,
such as the EUSO and OWL missions.  

The understanding of the high-energy universe is just beginning with the
promise of many new discoveries to come. By reading the messages brought
by all the different messengers we will gain important insight into the
workings of Nature and the origin and evolution of the Universe.

\begin{figure}
\begin{center}
\includegraphics[width=0.7\textwidth]{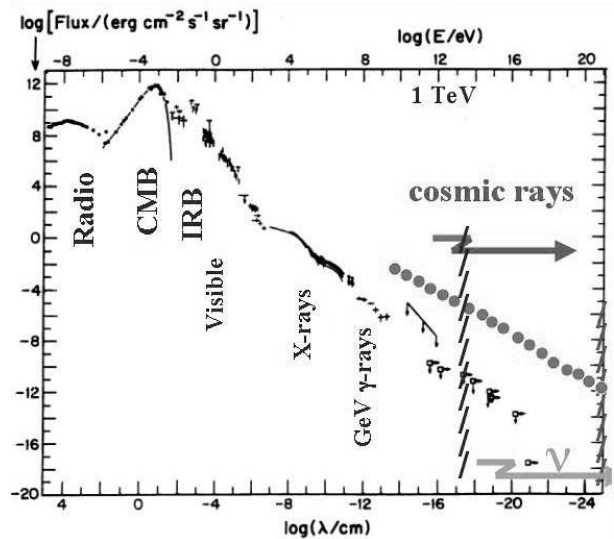}
\caption{ Spectra of photon backgrounds (based on Ressel and
Turner$^{11}$ modified by Halzen$^{3}$, cosmic rays, and
expected neutrino flux from ultra high energy cosmic ray interactions.}
\label{fig:il}
\end{center}
\end{figure}

\section*{Acknowledgments}
We thank the organizers of the XXI Texas Symposium. This
work was supported in part by the NSF through grant AST-0071235 and DOE
grant DE-FG0291-ER40606 at the University of Chicago.

\end{document}